\documentclass[review]{elsarticle}

\usepackage{siunitx}
\usepackage{mathtools}
\usepackage{soul}
\newcommand{\figref}[1]{Fig.~\ref{#1}}
\newcommand{\secref}[1]{Section~\ref{#1}}
\newcommand{\eqnref}[1]{Eq.~(\ref{#1})}
\newcommand{\tabref}[1]{Table~\ref{#1}}

\newcommand{\punc}[1]{\,#1}
\newcommand{\anglemean}[1]{\left\langle#1\right\rangle}
\newcommand{\change}[1]{#1}

\biboptions{numbers,sort&compress}

\newcommand{\arr}[2]{\begin{array}{#1}#2\end{array}}   

\newcommand{\vx}{\textbf{x}}
\newcommand{\vf}{\textbf{f}}

\newcommand{\margtodo}                                 
{\marginpar{\textbf{\textcolor{kitcolor}{ToDo}}}{}}
{\begin{color}{gray}}{\end{color}}                     

\begin{document}

\title{Materials data validation and imputation with an artificial neural network}
\author{P.C. Verpoort}
\address{University of Cambridge, J.J. Thomson Avenue, Cambridge, CB3 0HE, United Kingdom}
\author{P. MacDonald}
\address{Granta Design, 62 Clifton Road, Cambridge, CB1 7EG, United Kingdom}
\author{G.J. Conduit}
\address{University of Cambridge, J.J. Thomson Avenue, Cambridge, CB3 0HE, United Kingdom}
\date{\today}

\begin{abstract}
We apply an artificial neural network to model and verify material
properties. The neural network algorithm has a unique capability to handle
incomplete data sets in both training and predicting, \change{so it can
regard properties as inputs allowing it to} exploit \change{both}
composition-property and property-property correlations to enhance the
quality of predictions, and can also handle a graphical data as a single
entity. The framework is tested with different validation schemes, and then
applied to materials case studies of alloys and polymers. The algorithm
found twenty errors in a commercial materials database that were confirmed
against primary data sources.
\end{abstract}

\maketitle

\section{Introduction}

Through the stone, bronze, and iron ages the discovery of new materials has
chronicled human history. The coming of each age was sparked by the chance
discovery of a new material. However, materials discovery is not the only
challenge: selecting the correct material for a purpose is also
crucial\cite{Ashby04}. Materials databases curate and make available
properties of a vast range of
materials\cite{Jain13,NoMaD,MatWeb,MaterialUniverse,ProspectorPlastics}. However,
not all properties are known for all materials, and furthermore, not all
sources of data are consistent or correct, introducing errors into the data
set. To overcome these shortcomings we use an artificial neural network
(ANN) to uncover and correct errors in the commercially available database
\textit{MaterialUniverse}\cite{MaterialUniverse} and \textit{Prospector
  Plastics}\cite{ProspectorPlastics}.

Many approaches have been developed to understand and predict materials
properties, including direct experimental measurement\cite{Zhang08},
heuristic models, and first principles quantum mechanical
simulations\cite{Tadmor11}. We have developed an ANN algorithm that can be
trained from materials data to rapidly and robustly predict the properties
of unseen materials.\cite{conduit17} Our approach has a unique ability to
handle the data sets \change{that typically have incomplete data for input
  variables.}  \change{Such} incomplete \change{entries} would usually be
discarded, but the approach presented will exploit it to gain deeper
insights into material correlations. Furthermore, the tool can exploit the
correlations between different materials properties to enhance the quality
of predictions. The tool has previously been used to propose new optimal
alloys\cite{conduit17,Conduit2013ii,Conduit2013iii,Conduit2014v,Conduit2014vi,Conduit2014vii},
but here we use it to impute missing entries in a materials database and
search for erroneous entries.

Often, material properties cannot be represented by a single number, as they
are dependent on other test parameters such as temperature. They can be
considered as a graphical property, for example yield stress versus
temperature curves for different alloys\cite{Ritchie73}. In order to handle
this type of data more efficiently, we treat the data for these graphs as
vector quantities, and provide the ANN with information of that curve as a
whole when operating on other quantities during the training
process. \change{This requires less data to be stored than the typical
approach to regard each point of the graph as a new material, and allows a
generalized fitting procedure that is on the same footing as the rest of
the model.}

Our proposed framework is first tested and validated using generated
exemplar data, and afterwards applied to real-world examples from the
\textit{MaterialUniverse} and \textit{Prospector Plastics} databases. The
ANN is trained on both the alloys and polymers data sets, and then used to
make predictions to identify incorrect experimental measurements, which we
correct using primary source data. For materials with missing data entries,
for which the database provides estimates from modeling functions, we also
provide predictions, and observe that our ANN results offer an improvement
over the established modeling functions, while also being more robust and
requiring less manual configuration.

In \secref{sec:formalism} of this paper, we cover in detail the novel
framework that is used to develop the ANN. We compare our methodology to
other approaches, and develop the algorithms for computing the outputs from
the inputs, iteratively replacing missing entries, promoting graphing
quantities to become vectors, and the training
procedure. \secref{sec:Testing} focuses on validating the performance of the
ANN. The behavior as a function of the number of hidden nodes is
investigated, and a method of choosing the optimal number of hidden nodes is
presented. The capability of the network to identify erroneous data points
is explained, and a method to determine the number of erroneous points in a
data set is presented. The performance of the ANN for training and running
on incomplete data is validated, and tests with graphing data are
performed. \secref{sec:CaseStudies} applies the ANN to real-world examples,
where we train the ANN on \textit{MaterialUniverse}\cite{MaterialUniverse}
alloy and \textit{Prospector Plastics}\cite{ProspectorPlastics} polymer
databases, use the ANN's predictions to identify erroneous data, and
extrapolate from experimental data to impute missing entries.

\section{Framework}\label{sec:formalism}

Our knowledge of experimental properties of materials starts from a
database, a list of entries (from now on referred to as the `data set'),
where each entry corresponds to a certain material. Here, we take a property
to be either a defining property (such as the chemical formula, the
composition of an alloy, or heat treatment), or a physical property (such as
density, thermal conductivity, or yield strength)\cite{Ashby04}. The
following approach treats all of these properties on an equal footing.

To predict the properties of unseen materials a wide range of machine
learning techniques can be applied to such databases\cite{Oliveira16}.
Machine learning predicts based purely on the correlations between different
properties of the training data, which imbues the understanding of the
physical phenomena involved. We first define the ANN algorithm in
\secref{sec:ourframework}, and explain its implementation to incomplete data
in \secref{sec:frameworkfrag}.  Our extension to the ANN to account for
graphing data is described in \secref{sec:frameworkgraph}. The training
process is laid out in \secref{sec:frameworktrain}. Finally, we critically
compare our ANN approach to other algorithms in \secref{sec:comparetoGP}.

\subsection{Artificial Neural Network}\label{sec:ourframework}

\begin{figure}
 \centering
 \includegraphics[width=0.5\linewidth]{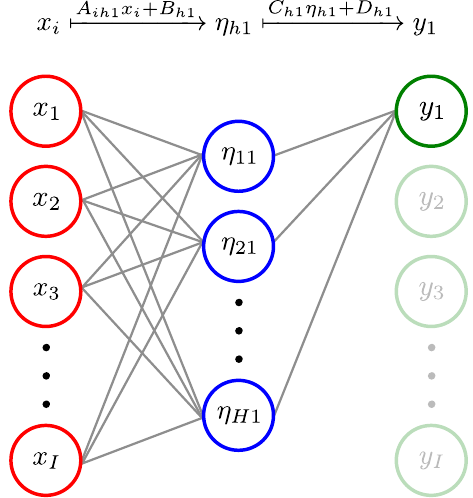}
 \\\vspace{0.5cm}
 \includegraphics[width=0.5\linewidth]{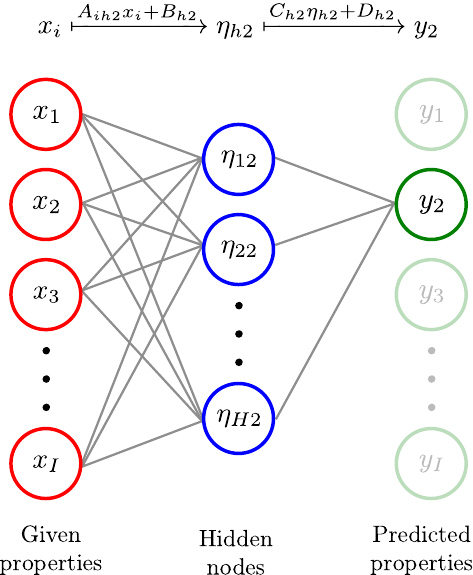}
 \caption{Artificial Neural Network (ANN) to interpolate data sets of
   materials properties. The graphs show how the outputs for $y_1$ (top) and 
   $y_2$ (bottom) are computed from all the inputs $x_i$. $I-2$ similar graphs 
   can be drawn for all other $y_j$ to compute all the predicted properties. 
   Linear combinations (gray lines on the left) of the
   given properties (red) are taken by the hidden nodes (blue), a non-linear
   tanh operation is applied, and a linear combination (gray lines on the
   right) of those is returned as the predicted property (green).}
 \label{fig:2-01a-sketch1}
\end{figure}

We now define the framework that is used to capture the functional relation
between all materials properties, and predict these relations for materials
for which no information is available in the data set. The approach builds
on the formalism used to design new nickel-base
superalloys~\cite{conduit17}. We intend to find a function $f$ that
satisfies the fixed-point equation $\vf(\vx)\equiv\vx$ as closely as
possible for all elements $\vx$ from the data set. There a total of $N$
entries in the data-set.  Each entry $\vx=(x_1,\ldots,x_{I})$ is a vector of
size $I$, and holds information about $I$ distinct properties.  The trivial
solution to the fixed-point equation is the identity operator, so that
$\vf(\vx)=\vx$. However, this solution does not allow us to use the function
$f$ to impute data, and so we seek a solution to the fixed-point equation
that by construction is orthogonal to the identity operator. This will allow
the function to predict a given component of $\vx$ from some or all other
components.

We choose a linear
superposition of hyperbolic tangents to model the function $\vf$,
\begin{align}
&\vf:(x_1,\ldots,x_i,\ldots,x_I)\mapsto(y_1,\ldots,y_j,\ldots,y_I)\label{eq:2-01-mapping}\\
&\arr{lrl}
{
 \qquad\mathrm{with}\qquad\qquad&
 y_j&=\sum_{h=1}^{H}C_{hj}\eta_{hj}+D_j,\\
 \qquad\mathrm{and} \qquad\qquad&
 \eta_{hj}&=\tanh\left(\sum_{i=1}^{I}A_{ihj}x_i+B_{hj}\right)\punc{.}
}
\nonumber
\end{align}
This is an ANN with one layer of hidden nodes, and is illustrated in
\figref{fig:2-01a-sketch1}. Each hidden node $\eta_{hj}$ with
$1\leq h\leq H$ and $1\leq j\leq I$ performs a $\tanh$ operation on a
superposition of input properties $x_i$ with parameters $A_{ihj}$ and
$B_{hj}$ for $1\leq i\leq I$. Each property is then predicted as a
superposition of all the hidden nodes with parameters $C_{hj}$ and
$D_j$. This is performed individually for each predicted property $y_j$ for
$1\leq j\leq I$. There are exactly as many given properties as predicted
properties, since all types of properties (defining and physical) are
treated equally by the ANN. Provided a set of parameters $A_{ihj}$,
$B_{hj}$, $C_{hj}$, and $D_j$, the predicted properties can be computed from
the given properties. The ANN always sets $A_{khk}=0$ for all
$1\leq k\leq I$ to ensure that the solution of the fixed-point equation is
orthogonal to the identity, and so we derive a network that can predict
$y_k$ without the knowledge of $x_k$.

\subsection{Handling incomplete data} \label{sec:frameworkfrag}

\begin{figure}
 \centering
 \includegraphics[width=0.5\linewidth]{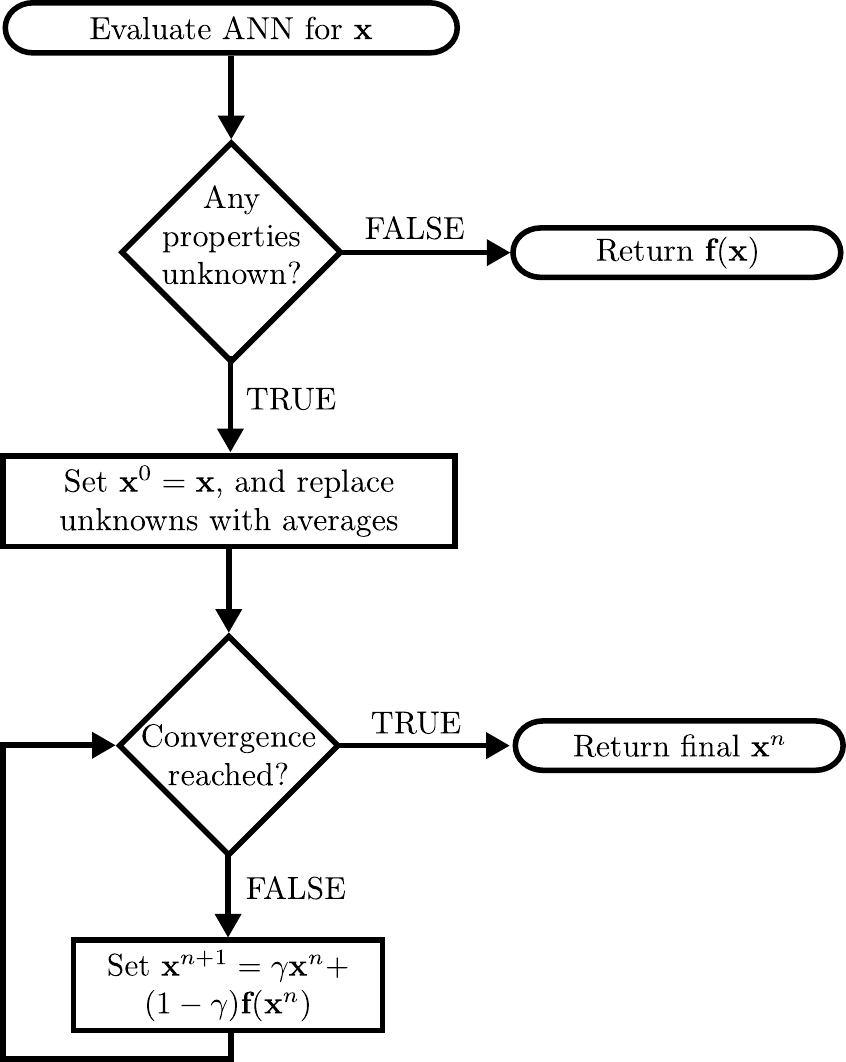}
 \caption{If we want to evaluate the ANN for a data point $\vx$ which has
   some of the entries for its properties missing, we will follow the
   process described by this graph. After checking for the trivial case
   where all entries are existent, we set $\vx^0 = \vx$, and replace all the
   missing entries by averages from the training data set. We then
   iteratively compute $\vx^{n+1}$ as a combination $\vx^n$ and $\vf$
   applied to $\vx^n$ until a certain point of convergence is reached, and
   return the final $\vx^n$ as a result instead of $\vf(\vx)$.}
 \label{fig:2-01b-sketch2}
\end{figure}

Typically, materials data that has been obtained from experiments is
incomplete, i.e. not all properties are known for every material, but the
set of missing properties is different for each entry. \change{However,
there is information embedded within property-property relationships: for
example ultimate tensile strength is three times hardness.} A typical ANN
formalism requires that each property is either an input or an output of the
network, and all inputs must be provided to obtain a valid
output. \change{In our example composition would be inputs, whereas ultimate
tensile strength and hardness are outputs. To exploit the known
relationship between ultimate tensile strength and hardness, and allow
either the hardness and ultimate tensile strength to inform missing data
in the other property}, we treat all properties as both inputs and outputs
of the ANN. \change{We have a single ANN rather than an exponentially large
number of them (one for each combination of available composition and
properties).}  We then adopt an expectation-maximization
algorithm\cite{Krishnan08}. This is an iterative approach, where we first
provide an estimate for the missing data, and then use the ANN to
iteratively correct that initial value.

The algorithm is shown in \figref{fig:2-01b-sketch2}. For any material $\vx$
we check which properties are unknown. In the non-trivial case of missing
entries, we first set missing values to the average of the values present in
the data set. An alternative approach would be to adopt a value suggested by
that of a local cluster. With estimates for all values of the neural network
we then iteratively compute
\begin{align}
 \vx^{n+1}=\gamma\vx^n+(1-\gamma)\vf(\vx^n)\punc{.}
\end{align}
The converged result is then returned instead of $\vf\textbf(\vx)$. The function $\vf$
remains fixed on each iteration of the cycle.

We include a softening parameter $0\le\gamma\le1$. With $\gamma=0$ we ignore
the initial guess for the unknowns in $\vx$ and determine them purely by
applying $\vf$ to those entries. However, introducing $\gamma>0$ will
prevent oscillations and divergences of the sequence, typically we set
$\gamma=0.5$.

\subsection{Functional properties} \label{sec:frameworkgraph}

Many material properties are functional graphs, for example to capture the
variation of the yield stress with temperature\cite{Ritchie73}. To handle
this data efficiently, we promote the two varying quantities to become
interdependent vectors. This will reduce the amount of memory space and
computation time used by a factor roughly proportional to the number of
entries in the vector quantities. \change{It also allows the tool to model
  functional properties on the same footing as the main model, rather than
  as a parameterization of the curve such as mean and gradient. The graph is
  represented by a series of points indexed by variable $\ell$.} Let $\vx$
be a point from a training data set. Let $x_{1,\ell}$ and $x_{2,\ell}$ be
the varying graphical properties, and let all other properties
$x_3,x_4,\ldots$ be normal scalar quantities. When $\vf(\vx)$ is computed,
the evaluation of the vector quantities is performed individually for each
component of the vector,
\begin{align}
 y_{1,l}=f_1(x_{1,\ell},x_{2,\ell},x_3,x_4,\ldots)\punc{.}
\end{align}
When evaluating the scalar quantities, we aim to provide the ANN with
information of the $x_2(x_1)$ dependency as a whole, instead of the
individual data points (i.e. parts of the vectors $x_{1,\ell}$, and
$x_{2,\ell}$). It is reasonable to describe the curve in terms of different
moments with respect to some basis functions for modeling the curve. For
most expansions, the moment that appears in lowest order is the average
$\anglemean{x_1}$, or $\anglemean{x_2}$ respectively. We therefore evaluate
the scalar quantities by computing,
\begin{align}
 y_3=f_3(\anglemean{x_1},\anglemean{x_2},x_3,x_4,\ldots)\punc{.}
\end{align}
This can be extended by defining a function basis for expansion, and 
include their higher order moments. \change{This approach automatically
  removes the bias due to differeing numbers of points in the graphs.}

\subsection{Training process}\label{sec:frameworktrain}

The ANN has to first be trained on a provided data set. Starting from random
values for $A_{ihj}$, $B_{hj}$, $C_{hj}$, and $D_j$, the parameters are
varied following a random walk, and the new values are accepted, if the new
function $\vf$ models the fixed-point equation $\vf(\vx)=\vx$ better. This
is quantitatively measured by the error function,
\begin{align}
 \delta=\sqrt{\frac{1}{N}\sum_{x\in X}\sum_{j=1}^I\left[ f_j(\vx)-x_j\right]^2}\punc{.}
 \label{eq:rms}
\end{align}
The optimization proceeds by a steepest descent approach\cite{Floudas08},
where the number of optimization cycles $C$ is a run-time variable.

In order to calculate the uncertainty in the ANN's prediction,
$\vf^\sigma (\vx)$, we train a whole suite of ANNs simultaneously, and
return their average as the overall prediction and their standard deviation
as the uncertainty\cite{Steck03}. We choose the number of models $M$ to be
between $4$ and $64$, since this should be sufficient to extract the mean
and uncertainty. In \secref{sec:Testing} we show how the uncertainty
reflects the noise in the training data and uncertainty in
interpolation. Moreover, on systems that are not uniquely defined, knowledge
of the full distribution of models will expose the degenerate solutions.

\subsection{Alternative approaches} \label{sec:comparetoGP}

ANNs like the one proposed in this paper (with one hidden layer and a
bounded transfer function; see \eqnref{eq:2-01-mapping}) can be expressed as
a Gaussian process using the construction first outlined by Neal
\cite{NEAL96} in 1996. Gaussian processes were considered as an alternative
to building the framework in this paper, but were rejected for two
reasons. Firstly, the ANNs have a lower computational cost, which scales
linearly with the number of entries $N$, and therefore ANNs are feasible to
train and run on large-scale databases. The cost for Gaussian processes
scales as $N^3$, and therefore does not provide the required
speed. Secondly, materials data tends to be clustered. Often, experimental
data is easy to produce in one region of the parameter space, and hard to
produce in another region. Gaussian processes can only define a unique
length-scale of correlation and consequently fail to model clustered data
whereas ANNs perform well.

\section{Testing and validation}\label{sec:Testing}

Having developed the ANN formalism, we proceed by testing it on exemplar
data. We will take data from a range of models to train the ANN, and
validate its results. We validate the ability of the ANN to capture
functional relations between materials properties, handle incomplete data,
and calculate graphical quantities.

In \secref{sec:BasicTests}, we interpolate a set of 1-dimensional functional
dependencies (cosine, logarithmic, quadratic), and present a method to
determine the optimal number of hidden nodes. In \secref{sec:ErroneousData},
we demonstrate how to determine erroneous entries in a data set, and to
predict the number of remaining erroneous
entries. \secref{sec:FragmentedData} provides an example of the ANN
performing on incomplete data sets. Finally, in \secref{sec:FunctionalData},
we present a test for the ANN's graphing capability.

\subsection{One-dimensional tests}\label{sec:BasicTests}

\begin{figure*}
 \centering
 \includegraphics[width=0.49\linewidth]{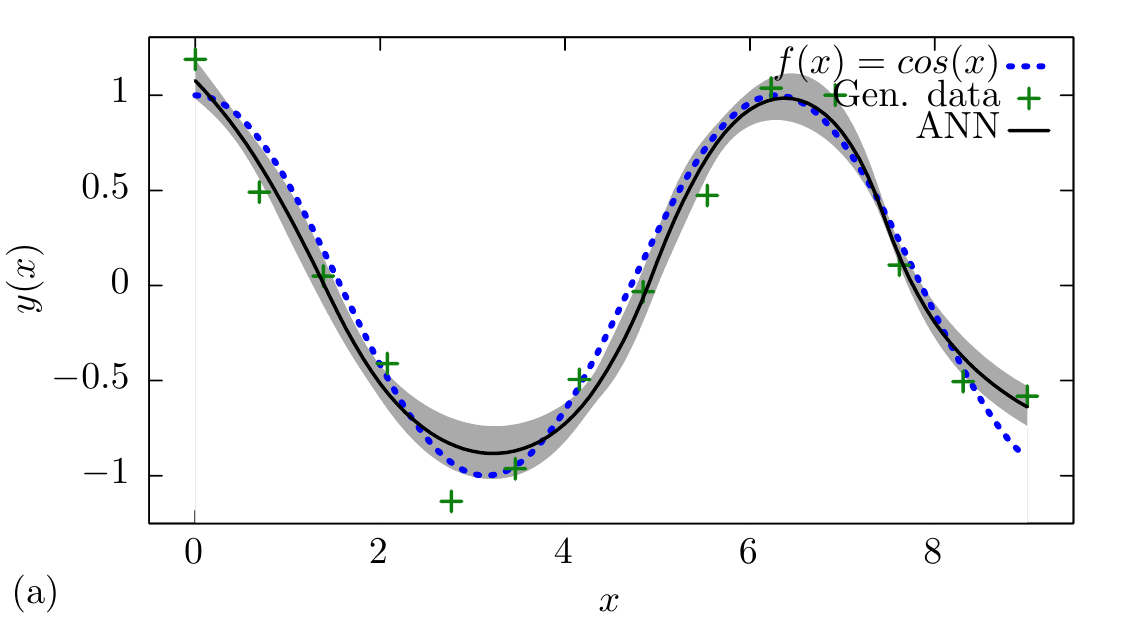}
 \includegraphics[width=0.49\linewidth]{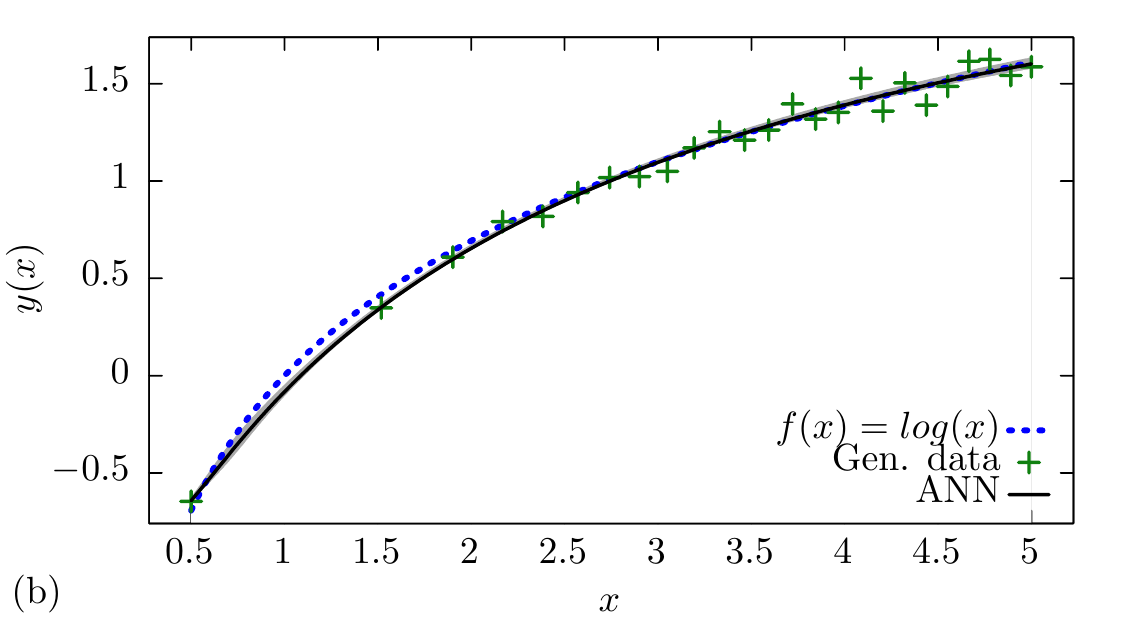}
 \includegraphics[width=0.49\linewidth]{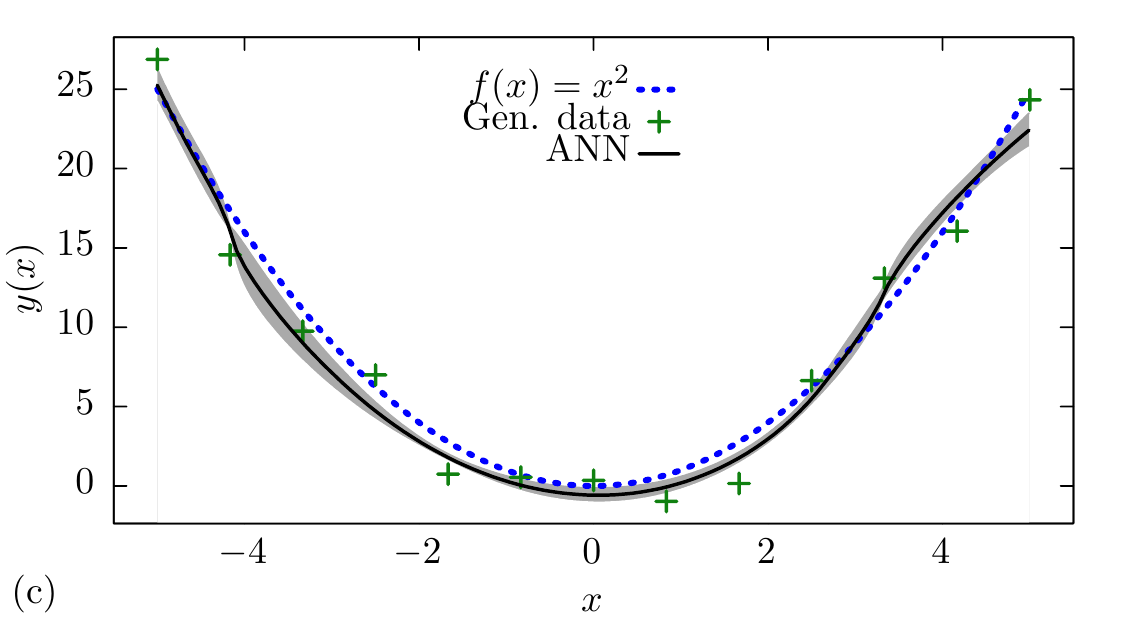}
 \includegraphics[width=0.49\linewidth]{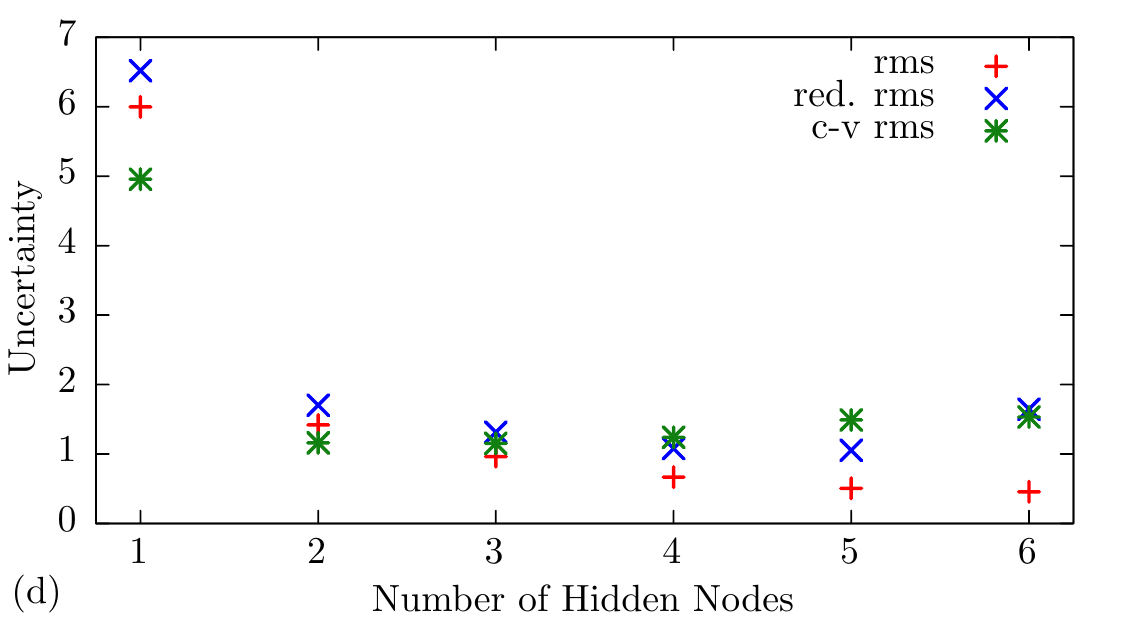}
 \caption{Training an ANN on toy-model data for (a) a cosine function, (b) a
   logarithmic function with unequally distributed data, and (c) a quadratic
   function with Gaussian noise. (d) For the quadratic function, the
   performance with different number of hidden nodes is tested, and the rms
   (\eqnref{eq:rms}), the reduced rms (\eqnref{eq:rmsred}), and the
   cross-validation rms are computed and plotted.}
 \label{fig:toymodeltest}
\end{figure*}

\begin{table}
  \caption{\change{The results of cross-validation testing for the three
      models (a) a cosine function, (b) a logarithmic function with
      unequally distributed data, and (c) a quadratic function with Gaussian
      noise. The second column gives the error when the ANN is trained on
      all of the data, and the third column the error in points unseen in
      training during cross-validation.}}
\begin{tabular}{lcc}
  \bf{Data set}&
  \bf{Error all}&
  \bf{Error cross-validation}\\
  \hline
  Cosine&\num{0.06}&\num{0.07}\\
  Logarithm&\num{0.05}&\num{0.06}\\
  Quadratic&\num{1.2}&\num{1.4}
\end{tabular}
  \label{tab:3-01-cross}
\end{table}

The ANN was trained on a (a) cosine function, (b) logarithmic function with
unequally distributed data, and (c) quadratic function with results shown in
\figref{fig:toymodeltest}. All of the data is generated with Gaussian
distributed noise to reflect experimental uncertainty in real-world material
databases. The cosine function is selected to test the ability to model a
function with multiple turning points, and was studied with $H=3$ hidden
nodes.  The logarithmic function is selected because it often occurs in
physical examples such as precipitate growth, and is performed with
$H=1$. The quadratic function is selected because it captures the two lowest
term in a Taylor expansion, and is performed with $H=2$.

\figref{fig:toymodeltest} shows that the ANN recovers the underlying
functional dependence of the data sets well. The uncertainty of the model is
larger at the boundaries, because the ANN has less information about the
gradient. The uncertainty also reflects the Gaussian noise in the training
data, as can be observed from the test with the $\log$ function, where we
increased the Gaussian noise of the generated data from left to right in
this test. For the test on the $\sin$ function, the ANN has a larger
uncertainty for maxima and minima, because these have higher curvature, and
are therefore harder to fit. \change{The correct modeling of the smooth
  curvature of the cosine curve could not be captured by simple linear
  interpolation.}

The choice of the number of hidden nodes $H$ is critical: Too few will
prevent the ANN from modeling the data accurately; too many hidden nodes
leads to over-fitting. To study the effect of changing the number of hidden
nodes, we repeat the training process for the quadratic function with
$1 \leq H \leq 6$, and determine the error $\delta$ in three ways. Firstly,
the straight error $\delta$. The second approach is cross-validation by
comparing to additional unseen data\cite{Hill05}. The third and final
approach is evaluate the reduced error
\begin{align}
\delta^*=\frac{\delta}{\sqrt{1-2H/N}}\punc{,}\label{eq:rmsred}
\end{align}
which assumes that the sum of the squares in \eqnref{eq:rms} is
$\chi^2$-distributed, so we calculate the error per degree of freedom, which
is $N-2H$, where the $2H$ parameters in the ANN arise because each of the
$H$ indicator functions in \eqnref{eq:2-01-mapping} has two degrees of
freedom: a scaling factor and also a shift. The results are presented in
\figref{fig:toymodeltest}(d).

The error, $\delta$, monotonically falls with more hidden nodes. This is
expected as more hidden nodes gives the model the flexibility to describe
the training data more accurately. However, it is important that the ANN
models the underlying functional dependence between those data points well,
and does not introduce overfitting.  The cross-validation results increase
above $H=2$ hidden nodes, which implies that overfitting is induced beyond
this point. Therefore, $H=2$ is the optimal number of hidden nodes for the
quadratic test. This is expected since we choose $\tanh$ as the basis
functions to build our ANN, which is a monotonic function, and the quadratic
consists of two parts that are decreasing and increasing respectively.

In theory, performing a cross-validation test may provide more insight into
the performance of the ANN on a given data set, however, this is usually not
possible because it has a high computational cost. We therefore turn to the
reduced error, $\delta^*$. This also has a minimum at $H=2$, and represents
a quick and robust approach to determine the optimal number of hidden nodes.

\change{Cross-validation also provides an approach to confirm the accuracy
  of the ANN predictions. For the optimal number of hidden nodes we perform
  a cross-validation analysis by taking the three examples in}
\figref{fig:toymodeltest}, \change{remove one quarter of the points at
  random, train a model on the remaining three quarters of the points, and
  then re-predict the unseen points. We then compare the error to the
  predictions of an ANN trained on the entire data set. The results are
  summarized in} \tabref{tab:3-01-cross}. \change{The error in the
  cross-validation analysis is only slightly larger than the error when
  trained off all entries, confirming the accuracy of the ANNs.}

In this section, we were able to prove that the ANN is able to model data
accurately, and laid out a clear prescription for determining the optimal
number of hidden nodes by minimizing $\delta^*$.

\subsection{Erroneous entries}\label{sec:ErroneousData}

\begin{figure}
 \centering
 \includegraphics[width=0.5\linewidth]{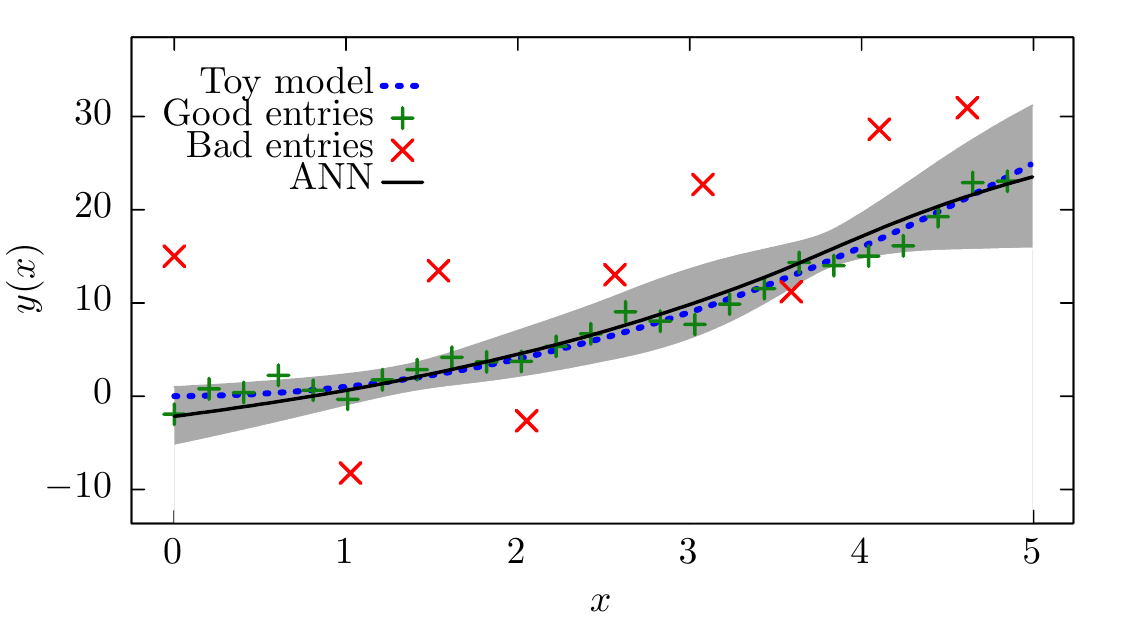}
 \caption{Blue dashed line: Quadratic curve for generating data. Red/green
   points: Data points generated from the blue line with Gaussian noise that
   have/have not been identified as erroneous. Black line: Prediction of the
   model with uncertainty. The Gaussian noise of the generated data
   increases proportional to the value of the toy-model function. Observe
   that less points are identified as erroneous at the right end of the
   plot, since the certainty of the ANN is lower in that region.}
 \label{fig:3C-01-plot}
\end{figure}

\begin{figure}
 \centering
 \includegraphics[width=0.5\linewidth]{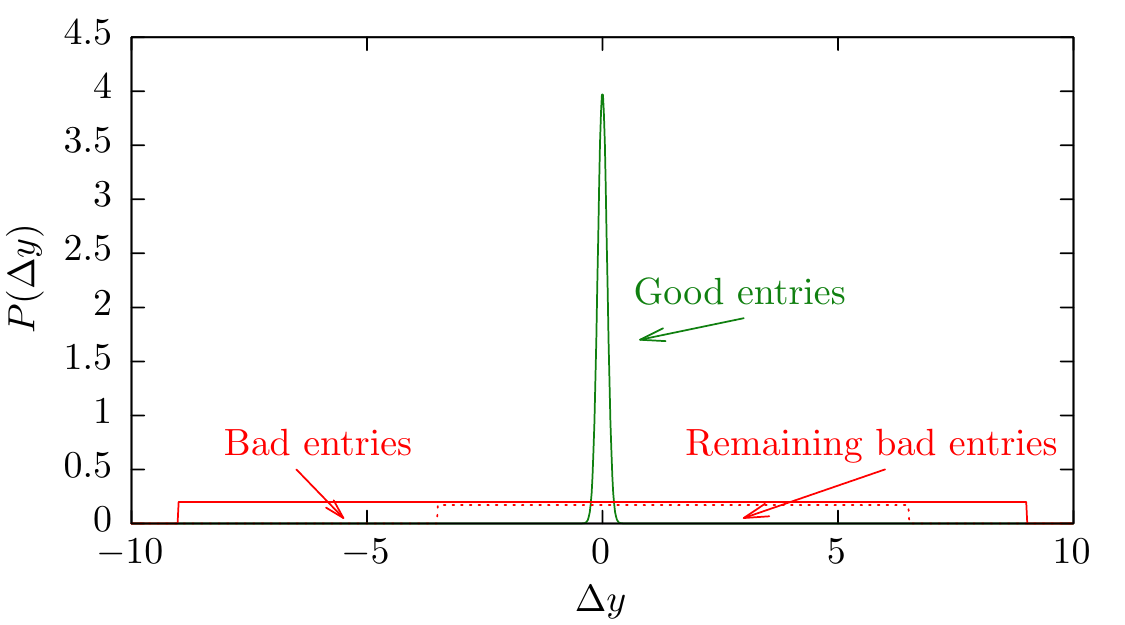}
 \caption{Theory of how many entries should be left using a uniform 
   distribution of 'bad' entries.}
 \label{fig:3C-02-remaining-theory}
\end{figure}

\begin{figure}
 \centering
 \includegraphics[width=0.5\linewidth]{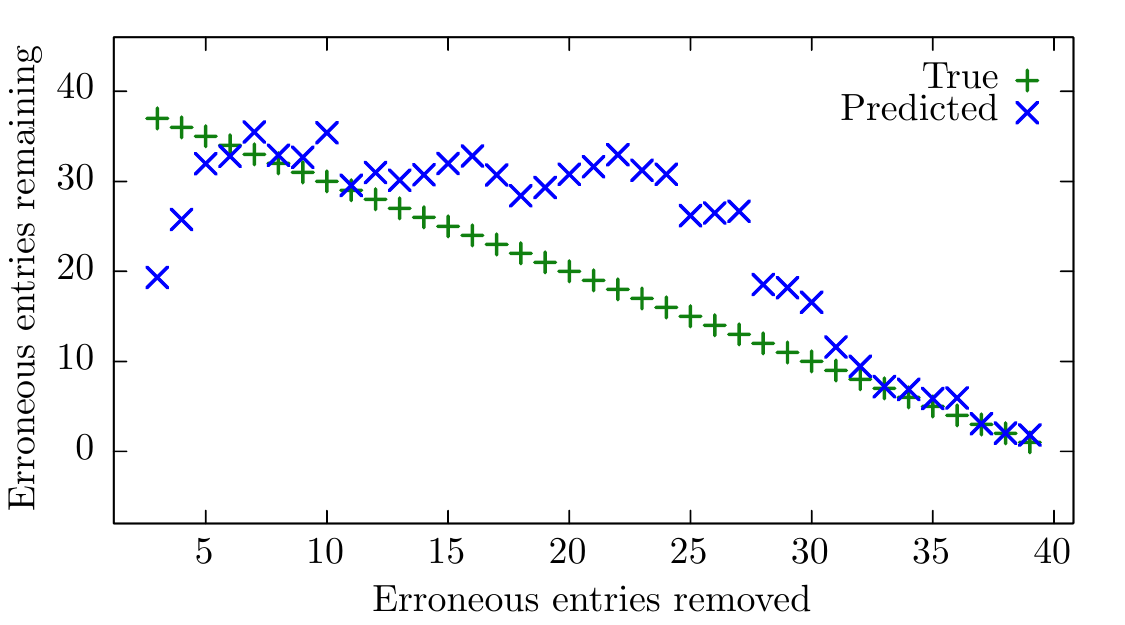}
 \caption{Blue points: the neural network prediction of, after removing a
   certain number of erroneous entries (x-axis), how many erroneous entries
   remain in the data set. A perfect prediction of the remaining number of
   erroneous entries would be the green points.}
 \label{fig:3C-03-remaining}
\end{figure}

The ANN can be used to search for erroneous entries in a data set. As the
ANN captures the functional dependence of the training data and the
uncertainty in the estimate, the likelihood of an entry being erroneous
can be determined by computing the number of standard deviations that this
entry lies away from the ANN's prediction,
\begin{align}
 \Delta_\sigma(\vx)=\sum_{j=1}^G\frac{f_j(\vx)-x_j}{f^\sigma_j(\vx)}\punc{.}
\end{align}
For a well-behaved data set with no errors the average absolute value of
$\Delta_{\sigma}$ should be approximately unity. However, in the presence of
erroneous entries, those entries with anomalously large $\Delta_\sigma(\vx)$
can be identified, removed, or corrected. In this section, we will analyze
the ability of the ANN to uncover erroneous entries in an exemplar set of
data.

The case study is based on a quadratic function shown in
\figref{fig:3C-01-plot} containing $N_\textrm{g}$ `good' points and
$N_\textrm{b}$ `bad' points. Good points would be the experimental data with
small Gaussian distributed noise, whereas bad points would occur through
strong systematic mistakes modeled with a broad uniform distribution
shown in \figref{fig:3C-02-remaining-theory}. The
results are shown in \figref{fig:3C-01-plot}, where only $\SI{25}{\percent}$
of the data is plotted.  The ten points that are identified to be the most
erroneous ones in this set are removed first, and have been highlighted
in the graph.

The upper limit of $\Delta_\sigma$ that we use to extract erroneous entries
from the data set has to be chosen correctly. We want to eliminate as many
erroneous entries as possible, while not removing any entries that hold
useful information.  We therefore proceed by developing a practical method
to analyze how many erroneous data entries are expected to remain in the
data set after extracting a certain number of entries. In a practical
application, the maintainer of a large materials database might opt to
continue removing erroneous entries from the database until the expected
number of erroneous entries that a user would encounter falls below 1.

The probability density for finding erroneous entries in the region where
erroneous entries have been removed from the sample is approximately equal
to the probability density for finding further erroneous entries in the
region of remaining entries. Therefore, the expected number of remaining
erroneous entries is
\begin{align}
 N_{\textrm{rem}}
 =\frac{N_{\textrm{found}}}{1-\frac{\Delta y_{\textrm{rem}}}{\Delta y_{\textrm{tot}}}}\punc{,}
 \label{eq:rempts}
\end{align}
where $N_\textrm{rem,found}$ are the number of remaining and found erroneous
data entries respectively, and $\Delta y_{\textrm{tot,rem}}$ refer to the
range over which the total and remaining entries are spread respectively.

Returning to the exemplar data set, we compare $N_{\textrm{rem}}$ with the true
number of remaining erroneous entries in \figref{fig:3C-03-remaining}. The
method provides a good prediction for the actual number of remaining
erroneous entries.

The neural network can identify the erroneous entries in a data
set. Furthermore, the tool can predict which are the entries most likely to
be erroneous allowing the end user to prioritize their attention on the
worse entries. The capability to predict the remaining number of entries
allows the end user to search through and correct erroneous entries until a
target quality threshold is attained.

\subsection{Incomplete data}\label{sec:FragmentedData}

In the following section, we investigate the capability of the ANN to train
on and analyze incomplete data. This requires at least three different
properties to study, and therefore our tests will be on three-dimensional
data sets. This procedure can be studied for different levels of correlation
between the properties, and we study two limiting classes: completely
uncorrelated, and completely correlated data. In the uncorrelated data set
the two input variables are uncorrelated with each other, but still
correlated to the output.  In the correlated data set the input variables are
now correlated with each other, and also correlated to the output. We focus
first on the uncorrelated data.

\subsubsection{Fully uncorrelated data}

\begin{figure}
 \centering
 \includegraphics[width=0.5\linewidth]{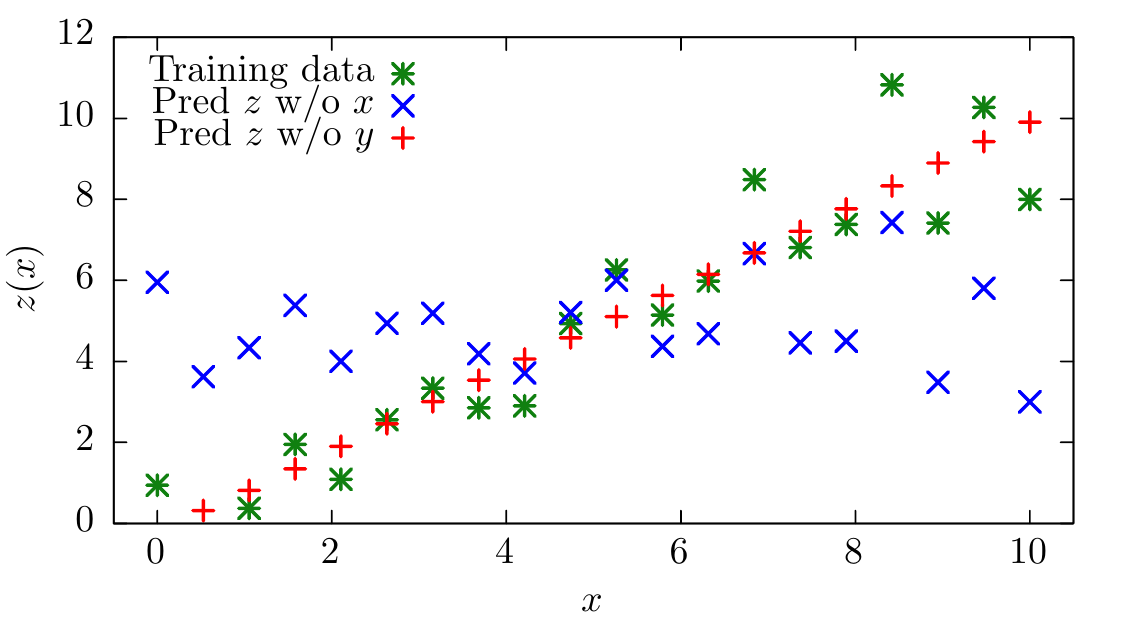}
 \caption{The toy-model data that is used for training of the ANN is shown
   as $Z$ as a function of $X$ together with the predictions of the ANN
   without providing $X$ or $Y$ respectively. The ANN learns the $Z=X+Y$
   dependence, and uses the average of $X$ or $Y$ values respectively to
   replace the unknown values.}
 \label{fig:3D-02-predictz-2}
\end{figure}

\begin{figure}
 \centering
 \includegraphics[width=0.5\linewidth]{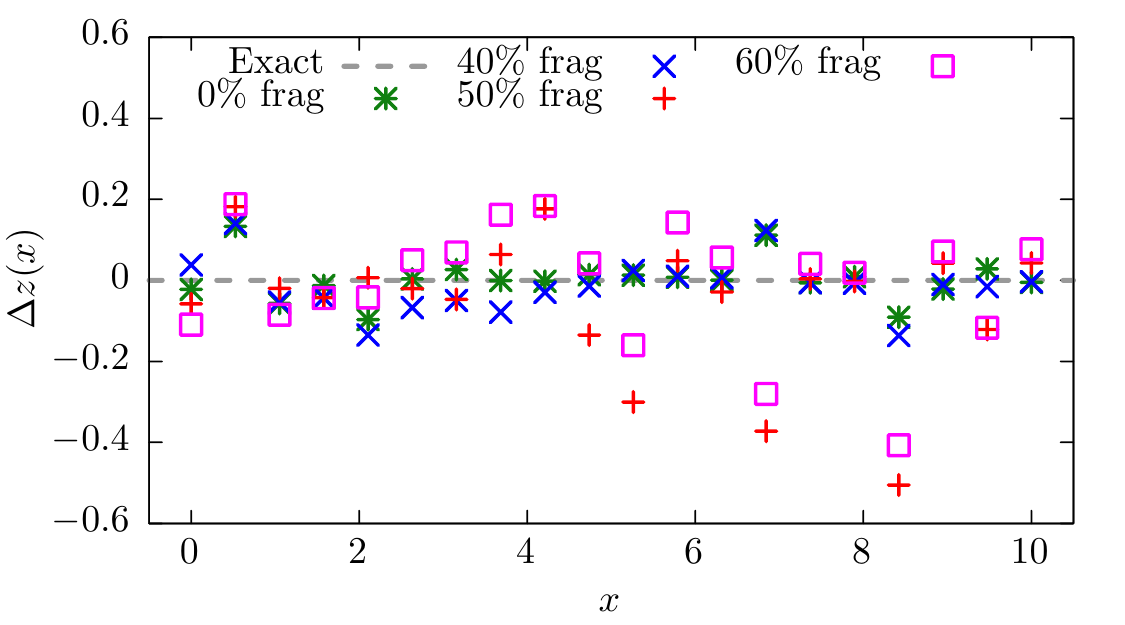}
 \caption{Deviation $\Delta z$ of the predicted values of $Z$ from the true
   values from the training data set as a function of $X$. An exact
   prediction would be represented by $\Delta z=0$ as indicated by the
   dashed gray line. The accuracy of the ANN predictions get worse with
   increasing level of fragmentation of the training data set.}
 \label{fig:3D-01-predictz-1}
\end{figure}

To study the performance on uncorrelated data we perform the following two
independent tests: we first train the ANN on incomplete uncorrelated data,
and run it on complete data, and secondly train on complete uncorrelated
data, and run on incomplete data.

For $N=20$ points $X=\{x_1,\ldots,x_N\}$ distributed evenly in the interval
$0\le x\le10$ we generate a set of random numbers $Y=(y_1,\ldots,y_N)$
uniformly distributed between \numrange{-2.5}{2.5}. We let
$Z=X+Y=(x_1+y_1,\ldots,x_N+z_N)$, which is shown in
\figref{fig:3D-02-predictz-2}. This data set is uncorrelated because the
values of $Y$, a set of random numbers, are independent of the values $X$;
therefore a model needs both $x$ and $y$ to calculate $z$.

We first train the ANN on all of the training data, and ask it to predict
$z$ while providing (i) $x$ and $y$, (ii) $x$ only, and (iii) $y$ only. The
results of (ii) and (iii) are shown in \figref{fig:3D-01-predictz-1}
alongside the training data, where $z$ is plotted as a function of $x$.
\figref{fig:3D-01-predictz-1} reveals that when provided with both $x$ and
$y$ the ANN is able to capture the full $z=x+y$ dependence for the complete
training data with maximum deviation $|\Delta z|\leq0.13$.  However, when
the ANN is provided only with the $x$ values, but not $y$, the best that the
ANN could do is replace $y$ with its average value,
$0$. \figref{fig:3D-02-predictz-2} confirms that the ANN returns
$z=x+\anglemean{Y}=x$. However, when the ANN is provided with the $y$ values
but not the $x$, the best that the ANN could do is replace $x$ with its
average value, $5$. \figref{fig:3D-02-predictz-2} shows that it returns
$z=\anglemean{X}+y=5+y$. This confirms that after training off a complete
uncorrelated data set, that when confronted with incomplete data, the ANN
delivers the best possible predictions given the data available.
\change{The analysis also confirms the behavior of the ANN when presented
  with completely randomly distributed data: it correctly predicts the mean
  value as the expected outcome.}

The second scenario is to train the ANN on an incomplete data set. Later,
when using the neural network to predict values of $z$, values for both $x$
and $y$ are provided.  We take the original training data, and randomly
choose a set of entries (in any of $X$, $Y$, or $Z$), and set them as
blank. We train the ANN on data sets that are (i) complete, (ii)
\SI{40}{\percent}, (iii) \SI{50}{\percent}, and (iv) \SI{60}{\percent}
missing values. The ANN is then asked to predict $z$ for given $x$ and $y$,
and the error in the predicted value of $z$ shown in
\figref{fig:3D-01-predictz-1}. The accuracy of the ANN predictions decreases
with increasing fragmentation of the training data. Yet, even with
\SI{60}{\percent} fragmentation the ANN is still able to capture the $z=x+y$
accurately with $\left|\Delta z\right|\leq0.41$ \SI{60}{\percent}. This is
less than the separation of $0.5$ between adjacent points in $Z$, so despite
over half of the data missing the ANN is still able to distinguish between
adjacent points.

\subsubsection{Fully correlated data}

We next turn to a data set in which given just one parameter, either $x$ or
$y$, it is possible to recover $z=x+y$. This requires that $y$ is a function
of $x$, and so is fully correlated. We now set $y=x^2$, and perform the
tests as above. Now after training on a complete data set, the ANN is able
to predict values for $z$ when given only $x$ or $y$. The ANN also performs
well when trained from an incomplete data set.

\subsubsection{Summary}

We have successfully tested the capability of the ANN to handle incomplete
data sets. We performed tests for both training and running the ANN with
incomplete data. The ANN performs well when the training data is both fully
correlated and completely uncorrelated, so should work well on real-life
data.

\subsection{Functional properties}\label{sec:FunctionalData}

\begin{figure}
 \centering
 \includegraphics[width=0.5\linewidth]{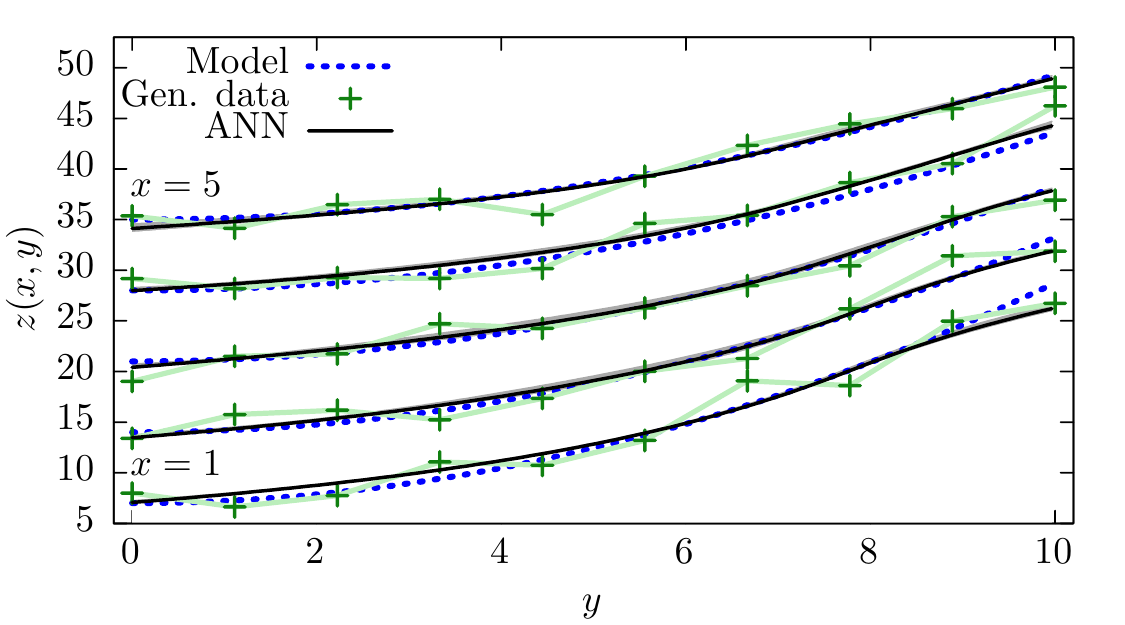}
 \caption{Training data, true function and predicted ANN function for
   different values of $x$.}
 \label{fig:3E-01-training-data}
\end{figure}

\begin{figure}
 \centering
 \includegraphics[width=0.5\linewidth]{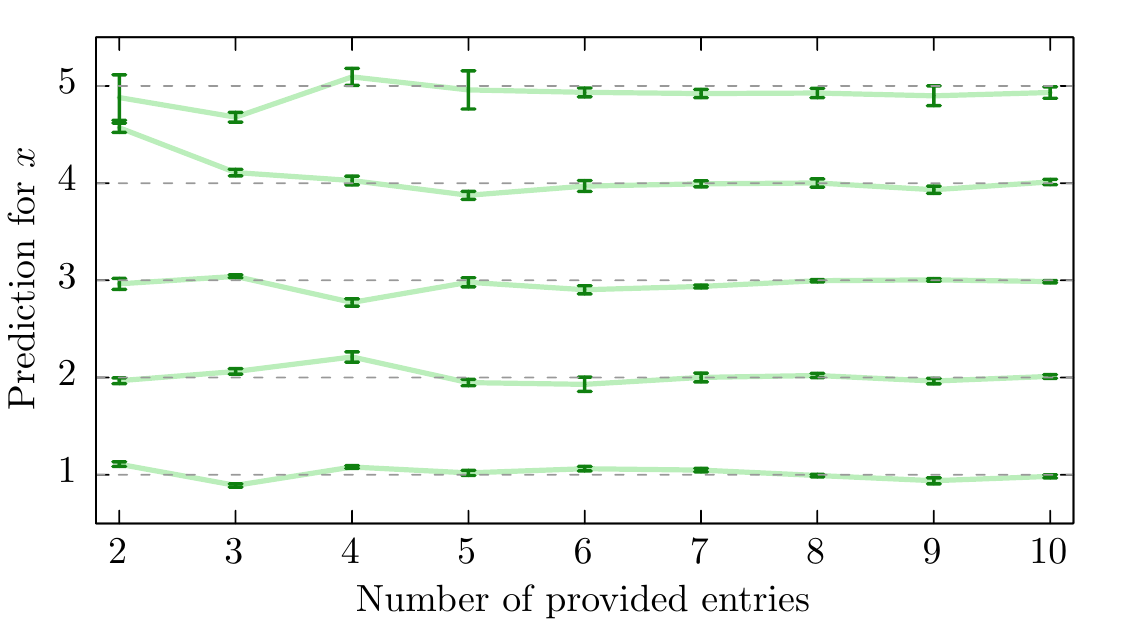}
 \caption{Predict $x$ from the training data with different number of $(y,z)$-
   pairs provided. The gray dotted lines indicate the true value of $x$.}
 \label{fig:3E-02-predict-x-numbofpts}
\end{figure}

We now test the capability for the ANN to handle data with functional
properties (also referred to as graphing data) as a single entity. As before
we have a functional variable $X=\{x_1,\ldots,x_N\}$ with $N=5$ equidistant
points in the interval from 1 to 5. At each value of $x$ point we introduce
a vector quantity, dependent on a variable $Y=\{y_{1},\ldots,y_{\ell}\}$
that can have up to $\ell=10$. We compute the vector function
$z_{\ell}=4x+y_{\ell}^2/4$, with additional Gaussian noise of width 1 and
train an ANN on this data set. We show the training data as well as the ANN
prediction for $z$ in \figref{fig:3E-01-training-data}, which confirms that
the ANN is able to predict the underlying functional dependence correctly.

The real power of the graphing capability is to predict $x$ with different
number of elements provided in the vector $(y,z)$. We show the predictions
of the ANN in \figref{fig:3E-02-predict-x-numbofpts}. With all 10 components
of the vector provided the ANN makes accurate predictions for $x$. With
fewer components provided the accuracy of the predictions for $x$ falls, but
even if just 2 elements are provided the ANN is still able to distinguish
between the discrete values of $x$.

We confirm that the ANN is able to fully handle vector and graphical
data. The ANN gives accurate predictions for both the functional properties
when providing non-functional properties only, and vice-versa. This new
capability allows the ANN to handle a new form of real-world problems, for
example the temperature dependence of variables such as yield
stress. Temperature can be used only as an input for the yield stress,
without the need to replicate other properties that are not temperature
dependent, for example cost. The reduction in the amount of data required
will increase the efficiency of the approach and therefore the quality of
the predictions.

\section{Applications}\label{sec:CaseStudies}

With the testing on model data complete we now present case studies of
applying the ANN to real-life data. In this section, we will use the ANN
framework to analyze the \textit{MaterialUniverse} and \textit{Prospector
  Plastics} databases. We first focus on a data set of 1641 metal alloys
with a composition space of 31 dimensions (that is each metal is an alloy of
potentially 31 chemical elements). We train neural networks of 4 hidden
nodes to predict properties such as the density, the melting point
temperature, the yield stress, and the fracture toughness of those
materials. Secondly, we examine a system where not all compositional
variables are available: a polymer data set of 5656 entries, and focus on
the modeling of its tensile modulus.

\change{We use the trained ANN to uncover errors by searching for
  entries multiple standard deviations $\Delta_{\sigma}$ away from the ANN
  predictions. We compare the results to primary sources referenced from the
  \textit{MaterialUniverse} data set to determine whether the entry was
  actually erroneous: a difference could only be due to a transcription
  error from that primary data set into the \textit{MaterialUniverse}
  database.}

When analyzing the density data, we can confirm the ability of the ANN to
identify erroneous entries with a fidelity of over \SI{50}{\percent}. For
the melting temperature data, we show that for missing entries the ANN
yields a significant improvement in the estimates provided by the curators
of the database. When advancing to the yield stress properties of the
materials, we observe that our methods can only be applied when additional
heat treatment data is made available for training the ANN. Unlike
established methods, our framework is uniquely positioned to include such
data for error-finding and extrapolation. For the fracture toughness data,
we exploit correlations with other known properties to provide more accurate
estimation functions compared to established ones.  Finally, in the polymer
data, we exploit the capability of our ANN to handle an incomplete data set
without compositional variables, and instead characterize polymers by their
properties.

\subsection{Density}\label{sec:density}

\begin{table}
  \caption{A list of \textit{MaterialUniverse} entries (source)
    for density in \si{\gram\per\cubic\centi\meter} that were
    identified by the ANN as being potentially 
    erroneous by the number of standard deviations $\Delta_{\sigma}$, and then
    subsequently confirmed to be incorrect by a primary source database (actual).}

\begin{tabular}{lcccc}
 \bf{Alloy}&
 \bf{Source}& 
 \bf{ANN}& 
 $\Delta_\sigma$&
 \bf{Actual}\\
 \hline
 Stainless steel, Ilium P&
 \num{7.6}&\num{7.9}&\num{12}&\numrange{7.75}{8.0}\cite{MatWeb}\\
 Tool steel, AISI M43& 
 \num{8.4}&\num{8.0}&\num{-12}&\numrange{7.7}{8.0}\cite{MatWeb}\\
 Copper-nickel, C70400&
 \num{8.5}&\num{8.9}&\num{11}&\num{8.9}\cite{MatWeb}\\
 Tool steel, AISI A3&
 \num{8.0}&\num{7.7}&\num{-20}&\num{8.9}\cite{MatWeb}\\
 Tool steel, AISI A4&
 \num{7.9}&\num{7.8}&\num{9}&\num{8.0}\cite{MatWeb}\\
 Tool steel, AISI M6&
 \num{8.5}&\num{8.0}&\num{11}&\numrange{7.7}{8.0}\cite{MatWeb}\\
 Aluminum, 8091, T6&
 \num{2.6}&\num{2.5}&\num{10}&\num{2.5}\cite{MatWeb}
\end{tabular}
  \label{tab:4-01-densities}
\end{table}

The density of an alloy is set primarily by its composition, the data for
which can be provided in a convenient form for training the ANN. This makes
the density data set an attractive starting point for our investigation.

We first construct a model following the rule of mixtures by calibrating a
weighted average of the densities of each constituent element to the
\textit{MaterialUniverse} density data. This model offers an average rms
error of \SI{0.19}{\gram\per\cubic\centi\meter}. We then construct a data
set that is the difference between the model and the original density data
and compositions, and use this to train the ANN. The resulting rms error in
the ANN prediction was \SI{0.12}{\gram\per\cubic\centi\meter}, a significant
improvement on the rule of mixtures.

With an ANN model for density in place, we use it to search for erroneous
entries within the density training set. For each entry we calculate the
number of standard deviations from the ANN prediction, with the top 20 being
considered as candidates for being erroneous. Of the 20 entries with highest
$\Delta_\sigma$, 7 were found to be incorrect after comparing to a primary
data source, these entries tabulated in \tabref{tab:4-01-densities}. Of the
remaining 13, 7 are found to be correct, and no source of primary data could
be found for the remaining 6. The ANN detected errors with a fidelity of
\SI{50}{\percent}. Following these amendments, using \eqnref{eq:rempts}, we
predict that the the number of remaining erroneous entries to be 17.

The ability to identify erroneous entries in a materials database, as well as
the ability to assess the overall quality should be of interest to the
curators of such databases. We therefore now use the ANN to search for
errors in other quantities in the \textit{MaterialUniverse} data set.

\subsection{Melting temperature}

\begin{table}
  \caption{Erroneous entries for melting temperature in \si{\kelvin}
    from the \textit{MaterialUniverse} database (source) alongside predictions from the ANN,
    that differ by $\Delta_{\sigma}$ standard deviations, subsequently confirmed
    to be incorrect by a primary source databases (actual).}
  \begin{tabular}{lcccc}
  \bf{Alloy}&
  \bf{Source}& 
  \bf{ANN}&
  $\Delta_\sigma$&
  \bf{Actual}\\
  \hline
  Wrought iron
  &\num{1973}&\num{1760}&\num{-37}&\num{1808}\cite{MatWeb}\\
  Nickel, INCOLOY840& 
  \num{1419}&\num{1661}&\num{8}&\numrange{1724}{1784}\cite{Incoloy840}\\
  Titanium, $\alpha$-$\beta$&   
  \num{1593}&\num{1878}&\num{17}&\num{1866}\cite{MatWeb}\\
  Steel, AISI 1095
  &\num{1650}&\num{1699}&\num{13}&\num{1788}\cite{AZOMaterials}
\end{tabular}
  \label{tab:4-02-temperatures}
\end{table}

\begin{table}
  \caption{Differences in the estimates of the melting temperature in
    \si{\kelvin} from the actual value for the 7 points where the
    established fitting function (EFF), and the ANN differ the most
    and primary source data is available.}
  \begin{tabular}{lcc}
  \bf{Alloy}& 
  $\Delta_\textrm{EFF}$& 
  $\Delta_\textrm{ANN}$\\ 
  \hline
  Steel Fe-9Ni-4Co-0.2C, quenched\cite{MetalSuppliersOnline}
  &-94&9\\
  Tool steel, AISI W5, water-hardened\cite{MetalSuppliersOnline}
  &48&-19\\
  Tool steel, AISI A4, air-hardened\cite{MetalSuppliersOnline}
  &56&16\\
  Tool steel, AISI A10, air-hardened\cite{AZOMaterials}
  &59&-61\\
  Tool steel, AISI L6, \SI{650}{\celsius} tempered\cite{MetalSuppliersOnline}
  &59&14\\
  Tool steel, AISI L6, annealed\cite{MetalSuppliersOnline}
  &59&14\\
  Tool steel, AISI L6, \SI{315}{\celsius} tempered\cite{MetalSuppliersOnline}
  &59&14
\end{tabular}
  \label{tab:4-08-temperatures-delta}
\end{table}

The melting point of a material is a complex function of its composition so
modeling it is a stern test for the ANN formalism. Furthermore, the melting
temperature data set in the \textit{MaterialUniverse} database has
\SI{80}{\percent} of its data taken from experiments with the remaining
\SI{20}{\percent} estimated from a fitting function by the database
curators. This means that we have to handle data with underlying differing
levels of accuracy.

We begin by training the ANN on only the experimental data. We seek to
improve the quality of the data set by searching and correcting erroneous
entries as was done for density. After identifying and correcting the 4
incorrect entries listed in \tabref{tab:4-02-temperatures}, we estimate that
there are still 5 erroneous entries in the data set. This leaves us with
just $\SI{0.3}{\percent}$ of the database being erroneous, and hence with a
high-quality data set of experimental measurements to study the accuracy of
the \textit{MaterialUniverse} fitting function.

We now wish to quantify the improvement in accuracy of the ANN model over
the established \textit{MaterialUniverse} fitting model for those entries
for which no experimental data is available. We do so by analyzing the 30
entries where the ANN and the fitting function are most different. By
referring to primary data sources we confirmed that the ANN predictions are
closer to the true value than the fitting function's prediction in 20 cases,
further away in 4 cases, and no conclusion is possible in 4 cases due to a
lack of primary data.

Sometimes there are two sources of primary data that are inconsistent.  In
these cases we can use the ANN to determine which source is
correct. Assuming that out of several experimental results only one can be
correct, we can decide which one it is by evaluating $\Delta_\sigma$ for
each entry, and comparing the resulting difference in likelihood for each of
the values being correct. For example, for the alloy \textit{AISI O1 Tool
  steel}, the value from one source is \SI{1694}{\kelvin}, only \num{0.6}
standard deviations away from the ANN prediction of \SI{1698}{\kelvin},
whereas the value given by the other source, \SI{1723}{\kelvin}, is
\num{4.5} standard deviations away.  The value of \SI{1694}{\kelvin}
$\exp{(-0.6^2)}/\exp{(-4.5^2)}\approx10^9$-times more likely to be correct
and we can therefore confidently adopt this value.

The ANN yields a clear improvement over the established fitting
model. Having accurate modeling functions available for is crucial for
operators of materials databases, and improvements over current modeling
functions will greatly benefit usage of those databases in industrial
applications.

\subsection{Yield stress}

\begin{table}
  \caption{The effect of adding heat treatment into the training set on the 
    average error in the ANN predictions of yield stress. Separate results for
    ferrous and non-ferrous alloys as well as the entire metals data set are shown.
    The error from the established fitting model used within
    \textit{MaterialsUniverse} is also shown.}
\begin{tabular}{lc}
  \bf{Data set}&
  \bf{Error}\\
  \hline
  Composition alone&\num{0.349}\\
  Composition and elongation&\num{0.092}\\
  Composition, elongation, and heat treatment&\num{0.052} \\
  Established model&\num{0.072}
\end{tabular}
 \label{tab:4-04-yield-stress-2}
\end{table}

\begin{table}
  \caption{Erroneous entries for yield stress in \si{\mega\pascal}
    from the \textit{MaterialUniverse} database (source) alongside
    predictions from the ANN, that differ by $\Delta_{\sigma}$
    standard deviations, subsequently confirmed
    to be incorrect by a primary source databases (actual).}
\begin{tabular}{lcccc}
  \bf{Alloy}&
  \bf{Source}& 
  \bf{ANN}&
  $\Delta_\sigma$& 
  \bf{Actual}\\
  \hline
  Stainless steel, AISI 301L& 
  \num{193}&\num{269}&\num{5}&\num{238}\cite{AZOMaterials}\\
  Stainless steel, AISI 301& 
  \num{193}&\num{267}&\num{5}&\num{221}\cite{AZOMaterials}\\
  Aluminum, 1080, H18& 
  \num{51} &\num{124}&\num{5}&\num{120}\cite{AZOMaterials}\\
  Aluminum, 5083, wrought& 
  \num{117}&\num{191}&\num{14}&\num{300},\num{190}\cite{MatWeb,AZOMaterials}\\
  Aluminum, 5086, wrought& 
  \num{110}&\num{172}&\num{11}&\num{269},\num{131}\cite{MatWeb,AZOMaterials}\\
  Aluminum, 5454, wrought& 
  \num{102}&\num{149}&\num{14}&\num{124}\cite{AZOMaterials}\\
  Aluminum, 5456, wrought& 
  \num{130}&\num{201}&\num{11}&\num{165}\cite{AZOMaterials}\\
  Nickel, INCONEL600& 
  \num{223}&\num{278}&\num{10}&\num{\ge 550}\cite{AZOMaterials}
\end{tabular}
 \label{tab:4-05-yield-stress-3}
\end{table}

We now study yield stress, a property of importance for many engineering
applications, and therefore one that must be recorded with high accuracy in
the \textit{MaterialUniverse} database. Yield stress is strongly influenced
by not only the composition but also the heat treatment routine. Initial
attempts to use composition alone produces an inaccurate ANN with relative
error of $0.349$ because alloys with similar or identical compositions had
undergone a different heat treatment and so have quite different yield
stress. To capture the consequences of the heat treatment routine additional
information can be included in the training set. For example, the elongation
depends on similar microscopic properties to yield stress, such as the bond
strength between atoms and the ease of dislocation movement, and so has a
weak inverse correlation with yield stress. Elongation was therefore
included in the training set, and as summarized in
\tabref{tab:4-04-yield-stress-2} we observed a reduction in the average
error to $0.092$ as a result.

To directly include information about the heat treatment a bit-wise
representation for encoding information on the range of different heat
treatments into input data readable by the ANN was devised. This was
achieved by representing the heat-treatment routine of an alloy
bit-wise, indicating whether or not the alloy had undergone the possible
heat treatments: tempering, annealing, wrought, hot or cold worked, or
cast. \tabref{tab:4-04-yield-stress-2} shows that including this heat
treatment data allows the ANN to model the data better than established
modeling frameworks, with the average error reduced to $0.052$. This error
can be compared with the standard polynomial fitting model previously used
by \textit{MaterialUniverse}, which has an error of $0.072$. This confirms
the increased accuracy offered by the ANN.

With the ANN model established, we can then use it to search for erroneous
entries within the \textit{MaterialUniverse} database. Following the
prescription developed in density and melting point, of the twenty alloys
with the largest $\Delta_\sigma$ in the estimate of yield stress, eight were
confirmed by comparison to primary sources to be erroneous, and are included
in \tabref{tab:4-05-yield-stress-3}. The other twelve entries could not be
checked against primary sources, resulting in an fidelity in catching errors
that could be confirmed of \SI{100}{\percent}.

\subsection{Fracture toughness}

\begin{table}
  \caption{Error in the ANN when different quantities are used 
    in the training data set to fit fracture toughness.}
\begin{tabular}{lc}
  \bf{Data set}&\bf{Relative error}\\
  \hline
  Composition alone&\num{0.144} \\
  Composition \& elongation& \num{0.113} \\
  Composition \& young modulus& \num{0.136} \\
  Composition \& yield stress& \num{0.132} \\
  Composition \& UTS&\num{0.134} \\
  Composition, elongation \& yield stress&\num{0.106}
\end{tabular}
  \label{tab:4-06-fracture-toughness-1}
\end{table}

\begin{table}
  \caption{Relative error in the available models for fracture 
    toughness, calculated over only the experimentally determined data.}
\begin{tabular}{lccccc}
  \bf{Model}&\bf{ANN}&\bf{Steels}&\bf{Nickel}&\bf{Aluminium}\\
  \hline
  Logarithmic error& \num{0.065} & \num{0.188} & 
  \num{0.102} & \num{0.086} \\
  Data points& \num{202} & \num{81} & \num{5} & \num{57}
\end{tabular}
  \label{tab:4-07-fracture-toughness-2}
\end{table}

Fracture toughness indicates how great a load a material containing a crack
can withstand before brittle fracture occurs. Although it is an important
quantity it has proven to be difficult to model from first principles. We
therefore turn to our ANN. Fracture toughness depends on both the stress
required to propagate a crack and the initial length of the crack. We can
therefore identify the UTS and yield stress as likely correlated quantities.
Additionally, elongation a measure of the material’s ability to deform
plastically, is also relevant for crack propagation.

The model functions fitted by the curator of \textit{MaterialUniverse} all
use composition as an input so we follow their prescription. An efficient
way to identify the properties most strongly correlated to fracture
toughness is to train the ANN with each quantity in turn (in addition to the
composition data), and then evaluate the deviation from the fracture
toughness data. The properties for which the error is minimized are the most
correlated.  \tabref{tab:4-06-fracture-toughness-1} shows that elongation is
the property most strongly correlated to fracture toughness. Whilst yield
stress, Young modulus, and UTS offer some reduction in the error, including
these quantities to the training data will not lead to a significant
improvement on the average error obtained from composition and elongation
alone.

The \textit{MaterialUniverse} fracture toughness data contains only around
$200$ values that have been determined experimentally, with the remaining
$1400$ values estimated by fitting functions. These are polynomial functions
which take composition and elongation as input, and are fitted to either
steels, nickel, or aluminum separately. We train the ANN over just the
experimentally determined data, and compare the error in its predictions
to those from the known fitting
functions. \tabref{tab:4-07-fracture-toughness-2} shows that the ANN is the
most accurate, having a smaller error than the fitting function for all
three alloy families. While the different fitting functions are `trained'
only on the subset of the data for which they are designed, the ANN is able
to use information gathered from the entire data set to produce a better
model over each individual subset. This is one of the key advantages of a
ANN over traditional data modeling methods.

\subsection{Polymers}

\begin{figure}
 \centering
 \includegraphics[width=0.5\linewidth]{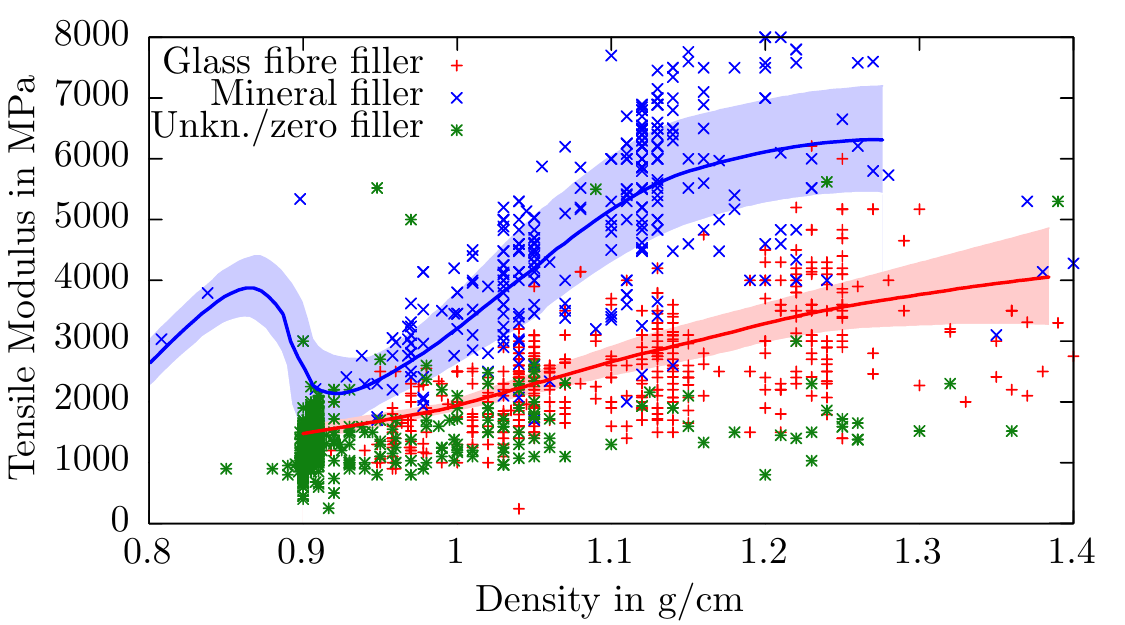}
 \caption{Polymer tensile modulus against density with glass fiber filler
   (blue) and mineral filler (red). The input information includes not only
   the filler type but also the filler amount (weight in \%).}
 \label{fig:4E-01-polymers}
\end{figure}

\begin{table}
  \caption{A list of three \textit{ProspectorPlastics} entries (source)
    for polymer flexural modulus/strength in \si{\MPa} that were
    identified by the ANN as being potentially erroneous and then
    subsequently confirmed to be incorrect by a  primary source
    databases (actual). The final five entries had missing filler 
    type and amount in $\%$ that imputed by the ANN and then confirmed
    against a primary data source.}
\begin{tabular}{lcccc}
 \bf{Polymer}&
 \bf{Property}&
 \bf{Source}& 
 \bf{ANN}&
 \bf{Actual}\\
 \hline
 4PROP25C21120&Modulus&
 \num{2300000}&\num{2186}&\num{2300}\cite{polyone}\\
 AZDELU400-B01N&Modulus&
 \num{8000000}&\num{8189}&\num{8000}\cite{polyone}\\
 Hyundai HT340&Strength&
 \num{469}&\num{46.1}&\num{46.9}\cite{polyone}\\
 Borealis NJ201AI&Mineral filler&
 -&$20\pm4$&\num{20}\cite{polyone}\\
 Daplen EE168AIB&Mineral filler&
 -&$11\pm3$&\num{10}\cite{polyone}\\
 Maxxam NM-818&Glass filler&
 -&$18\pm4$&\num{20}\cite{polyone}\\
 FORMULA P 5220&Mineral filler&
 -&$19\pm3$&\num{20}\cite{polyone}\\
 4PROP 9C13100&Mineral filler&
 -&$13\pm3$&\num{10}\cite{polyone}
\end{tabular}
  \label{tab:4-09-polymers}
\end{table}

In this section, we study polymers, which is an incomplete data set. Polymer
composition cannot be described simply by percentage of constituent elements
(as in the previous example with metals) due to the complexity of chemical
bonding, so we must characterize the polymers by their properties.  Some
properties are physical, such as tensile modulus and density; others take
discrete values, such as type of polymer or filler used, and filler
percentage. As the data~\cite{ProspectorPlastics} was originally compiled
from manufacturer's data sheets, not all entries for these properties are
known, rendering the data set incomplete.

We analyze a database of polymers that has the filler type as a
class-separable property. Many other properties exhibit a split based on
filler type, such as tensile modulus, flexural strength, or heat deflection
temperature. We first focus on the tensile modulus shown in
\figref{fig:4E-01-polymers}. Analysis of the predicted entries in
\tabref{tab:4-09-polymers} uncovers three erroneous entries that could be
confirmed against primary source data.  All three of these polymers had been
entered into the database incorrectly, being either one or three orders of
magnitude too large.

The data set is incomplete so many polymers have unknown filler type. The
vast majority of entries sit at the expected density of
\SI{0.9}{\gram\per\centi\meter\cubed}. However, some entries sit well away
from there. Since the data set includes no other properties that can account
for this discrepancy, a reasonable assumption is that these entries do not
have zero filler, but instead are lacking filler information. The ANN
applies this observation to predict the filler type and fraction. In
\tabref{tab:4-09-polymers} we show five polymers for which the filler type
and fraction were correctly predicted when compared to primary sources of
data.

Having successfully confirmed the ANN's ability to model incomplete data, we
have completed our tests on real-life data. The ANN can perform materials
data analysis that has so far not been possible with established methods,
and hence our framework yields an important improvement in operating
large-scale materials databases. With polymers being another class of
materials of great importance to industry, we have again shown how our
approach will have an impact across a broad range industrial fields.

\section{Conclusions}

We developed an artificial intelligence algorithm and extended it to handle
incomplete data, functional data, and to quantify the accuracy of data. We
validated its performance for model data to confirm that the framework
delivers the expected results in tests on the error-prediction, incomplete
data, and graphing capabilities. Finally, we applied the framework to the
real-life \textit{MaterialUniverse} and \textit{Prospector Plastics}
databases, and were able to showcase the immense utility of the approach.

In particular, we were able to propose and verify erroneous entries, provide
improvements in extrapolations to give estimates for unknowns, impute
missing data on materials composition and fabrication, and also help the
characterization of materials by identifying non-obvious descriptors across
a broad range of different applications. Therefore, we were able to show how
artificial intelligence algorithms can contribute significantly to
innovation in researching, designing, and selecting materials for industrial
applications.

The authors thank Bryce Conduit, Patrick Coulter, Richard Gibbens, Alfred Ireland, Victor
Kouzmanov, Hauke Neitzel, Diego Oliveira S\'anchez, and Howard Stone for
useful discussions, and acknowledge the financial support of the EPSRC
[EP/J017639/1] and the Royal Society. There is Open Access to this paper
and data available at \texttt{https://www.openaccess.cam.ac.uk}.

\end{document}